\documentclass[english,superscriptaddress]{revtex4-2}
\usepackage[utf8]{inputenc}
\usepackage{geometry}
\usepackage{xcolor}
\usepackage{babel}
\usepackage{float}
\usepackage{amsmath}
\usepackage{amssymb}
\usepackage{graphicx}
\usepackage[unicode=true,pdfusetitle,
 bookmarks=true,bookmarksnumbered=false,bookmarksopen=false,
 breaklinks=false,pdfborder={0 0 1},backref=false,colorlinks=false]
 {hyperref}

\makeatletter

\DeclareTextSymbolDefault{\textquotedbl}{T1}
\providecommand{\tabularnewline}{\\}

\usepackage{times}

\makeatother

\begin{document}
\title{Computing Shor's algorithmic steps with interference patterns of classical
light}
\author{Wei Wang}
\affiliation{Institute of Applied Physics and Materials Engineering, University
of Macau, Macau S.A.R., China}
\author{Ziyang You}
\affiliation{Institute of Applied Physics and Materials Engineering, University
of Macau, Macau S.A.R., China}
\author{Shuangpeng Wang}
\affiliation{Institute of Applied Physics and Materials Engineering, University
of Macau, Macau S.A.R., China}
\author{Zikang Tang}
\affiliation{Institute of Applied Physics and Materials Engineering, University
of Macau, Macau S.A.R., China}
\author{Hou Ian}
\email{houian@um.edu.mo}

\affiliation{Institute of Applied Physics and Materials Engineering, University
of Macau, Macau S.A.R., China}
\begin{abstract}
When considered as orthogonal bases in distinct vector spaces, the
unit vectors of polarization directions and the Laguerre-Gaussian
modes of polarization amplitude are inseparable, constituting a so-called
classical entangled light beam. Equating this classical entanglement
to quantum entanglement necessary for computing purpose, we show that
the parallelism featured in Shor's factoring algorithm is equivalent
to the concurrent light-path propagation of an entangled beam or pulse
train. A gedanken experiment is proposed for executing the key algorithmic
steps of modular exponentiation and Fourier transform on a target
integer $N$ using only classical manipulations on the amplitudes
and polarization directions. The multiplicative order associated with
the sought-after integer factors is identified through a four-hole
diffraction interference from sources obtained from the entangled
beam profile. The unique mapping from the fringe patterns to the computed
order is demonstrated through simulations for the case $N=15$.
\end{abstract}
\maketitle

\section{Introduction}

Non-zero correlation between space-like events was initially debated
in the famous Einstein-Podolsky-Rosen thought experiment~\citep{Einstein}.
This contemplated correlation was later quantified by Bell's inequality~\citealp{Bell}
and CHSH inequality~\citep{Clauser}, which subsequently give rise
to the concept of quantum entanglement~\citep{Mermin}. Violating
the law of locality, entanglement has long been thought of as a unique
feature only for systems operating in the quantum regime~\citep{Horodecki,Pang,Hobson}.
But recent investigations have revealed that entanglement to some
partial extent can be emulated in classical systems~\citep{Spreeuw,Qian,Aiello}.
In particular, the classical entanglement between the polarization
direction and the polarization amplitude is experimentally verified~\citep{Qian-1};
so is that between the polarization and the Orbital Angular Momentum
(OAM) in a Laguerre-Gaussian (LG) beam~\citep{Song}.

Entanglement plays a key role in implementing quantum algorithms.
The celebrated Shor's algorithm relies on the entanglement of two
quantum states, guised as two computational registers, to speed up
the finding of the correct factor of a large integer~\citep{Shor}.
The two registers can be physically realized by superconducting qubits~\citep{Lucero},
nuclear spins~\citep{Vandersypen}, or optical photons~\citep{O=002019Brien,Lu,Lanyon,Politi,Martin-lopez}.
Compared to quantum entanglement exemplified in these quantum systems,
classical entanglement obtains a similar correlation level while having
additional merits in coherence, generation, and detection~\citep{Forbes,Ladd},
making classical entangled systems substitute the roles of quantum
entangled states in communication and computation~\citep{T=0000F6ppel,Spreeuw-1,Deutsch}.
In this work, we apply the concept of classical entanglement to propose
a gedanken experiment using entirely classical sources and optical
elements to study the realizability of Shor's algorithm.

The existence of classical implementations of Deutsch algorithm~\citep{Perez-Garcia},
quantum walk~\citep{Goyal}, and quantum Fourier transform (QFT)~\citep{Song}
shows that quantum algorithms do not necessarily require a quantum
system to be realized. Rather, the border line that divides quantum
mechanics from classical mechanics does not exactly coincide with
the boundary between quantum and classical algorithms. The fully quantum
optical versions of Shor's algorithms~\citep{O=002019Brien} generate
entangled pairs of photons, which are let undergo probabilistically
selective paths with computational significance before one photon
is projectively measured and its entangled counterpart collapses into
the desired state. Here, we make use of the intrinsically entangled
degrees of freedom in the OAM amplitude modes and the polarization
directions of an LG beam to demonstrate an equivalent implementation
of the computational steps used by Shor's algorithm. The beam carries
out the parallel computational process by running a set of split beams
concurrently through multiple viable light paths before they are congregated
to generate a unique interference pattern~\citep{You} mappable to
the computation result. From this classical perspective, an equivalence
between the algorithmic complexity and the optical-path complexity
is established.

In particular, the OAM modes are used to represent the computational
basis of the control register while the polarization directions those
of the work register. The latter would store the data computed from
modular exponentiation and discrete Fourier transform, the two key
steps of Shor's algorithm. After the two-step process, a four-hole
interference setup sources from the unified beam profile a distinguishable
fringe pattern that has a unique correspondence to a specific entangled
computational bases in the two registers. We demonstrate this distinguishabililty
below of distinct multiplticative orders from fringe patterns for
the illustrative case of $N=15$ where only 4 OAM modes are needed.
To scale up the applicability for a larger $N$, we show that the
same setup can be used with a continuous-wave laser source replaced
by a pulse train. In the last part, we prove that the complexity of
our method is consistent with the original Shor's algorithm.

In the following, we present the algorithmic steps based on LG beams
in Sec.~\ref{sec:algorithm}. The optical path designs are explained
in Sec.~\ref{sec:optical_paths} and the detection methods discussed
in Sec.~\ref{sec:detection_method}. Sec.~\ref{sec:Complexity-analysis}
analyzes the complexity. Conclusions are given in Sec.~\ref{sec:conclusions}. 

\section{Classical entanglement and shor's algorithm~\label{sec:algorithm} }

Classical entanglement refers to non-separable correlations among
the degrees of freedom in classical optics. For instance, the two
orthogonal polarization directions ($\mathbf{e_{x}}$ and $\mathbf{e_{y}}$)
and their respective field amplitudes ($E_{x}$ and $E_{y}$) can
be taken as an entangled pair in the polarized optical field $\mathbf{E}=E_{x}\mathbf{e_{x}}+E_{y}\mathbf{e_{y}}$.
Allowing the field amplitudes to vary spatially beyond those depicted
by plane waves, one arrives at linearly polarized LG beams

\begin{equation}
\mathbf{E}=u_{p\text{,}l}\mathbf{H}+u_{p\text{,}-l}\mathbf{V},\label{eq: POL-OAM entanglement}
\end{equation}
where $\mathbf{H}$ and $\mathbf{V}$ denote the unit vectors for
respectively horizontal and vertical polarization directions while 

\begin{align}
u_{p\text{,}l}\left(r,\varphi,z\right) & =\frac{C(r\sqrt{2}/w_{z})^{\left|l\right|}}{\sqrt{1+z^{2}/z_{R}^{2}}}e^{-r^{2}/w_{z}^{2}}L_{p}^{\left|l\right|}\left(\frac{2r^{2}}{w_{z}^{2}}\right)\nonumber \\
\times & e^{-il\varphi}\exp i\left\{ \frac{kr^{2}z}{2(z^{2}+z_{R}^{2})}+(2p+\left|l\right|+1)\tan^{-1}\frac{z}{z_{R}}\right\} \label{eq: LG equation}
\end{align}
indicate the distribution function for the OAM-mode amplitudes in
cylindrical coordinates~\citep{Allen} with $z$ being the propagation
direction. In the equation, $C$ is a normalized constant, $z_{R}$
the Rayleigh range, $w_{z}$ the beam waist, and $L_{p}^{\left|l\right|}$
the associated Laguerre polynomial. It is worth noting that the function
has an azimuthal angular dependence of $\exp(-il\varphi)$, where
$l$ is the topological charge determining the angular momentum ($l\hbar$
per photon) and the chirality of helical phase fronts (sign of $l$). 

The set of OAM-modes specified by the integer $l$ forms an orthogonal
set of basis functions for a complete Hilbert space, i.e. the inner
product $(u_{p,l},u_{p,l'})=\delta_{l,l'}$. Hence, this OAM space
is mathematically isomorphic to a finite dimensional quantum state
space and we can use Dirac bra-ket notation for state $\left|l\right\rangle $
to indicate the OAM mode $u_{p,l}$. Similarly, the two polarization
directions are orthogonal, which forms a two-dimensional space independent
from the OAM modes and prompts us to use $\left|H\right\rangle $
and $\left|V\right\rangle $ to denote $\mathbf{H}$ and $\mathbf{V}$.
We make use of these orthogonalities to encode the computational bases
of two registers, a control register $\left|\psi_{C}\right\rangle $
and a work register $\left|\psi_{W}\right\rangle $, that Shor's algorithm
calls for.

After initialization, the joint computer state is separable, i.e.$\left|\psi_{C}\right\rangle \otimes\left|\psi_{W}\right\rangle $.
The first algorithmic step~\citep{Shor2} is the formation of the
entangled state

\begin{equation}
\frac{1}{\sqrt{2^{n}}}\sum_{x=0}^{2^{n}-1}\mathop{\left|x\right\rangle \left|a^{x}\textrm{mod}N\right\rangle }\label{eq: Shor entanglement}
\end{equation}
amongst the two registers, where the work register encodes the results
of modular exponentiation function (MEF) according to the exponent
$x$ given in the control register and the modulus $N$. The integer
$N$ is the target integer to be factored and the base $a<N$ is a
pre-selected number coprime with $N$, i.e. $\gcd(a,N)=1$. Number-theoretically
speaking, factorizing $N$ corresponds to finding an $x$ value such
that it becomes the multiplicative order of the MEF. The candidate
of this order $x$ lies within the range $\{0,1,\dots,2^{n}-1\}$,
hence the range of the summation, where $n$ denotes the number of
qubits necessary for constructing each register and is determined
by $N$, i.e. $2^{n}\approx\sqrt{N}$. Taking $N=15$ as an example,
one would have $n=2$ and $a=11$, making $x\in\{0,1,2,3\}$ while
the exponentiation assumes two possible values, $1$ and $11$.

\begin{figure}[H]
\noindent \centering{}\includegraphics[bb=0bp 0bp 393bp 135bp,clip,width=10cm]{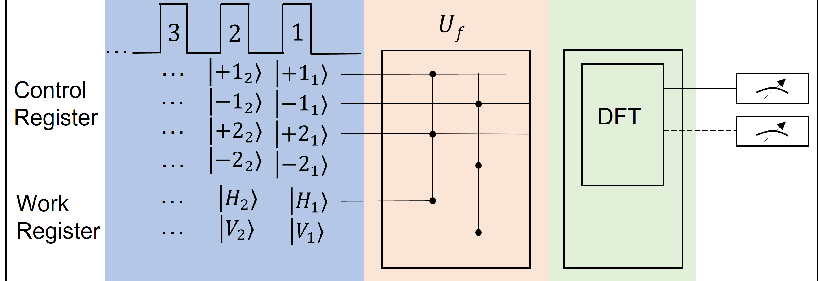}\caption{Circuit flow diagram for implementing Shor's algorithm. The blue,
pink, and green parts indicate the stages for initialization, modular
exponentiation, and discrete Fourier transform, respectively. The
blue arrow in MEF in this part refers to the polarization flip. The
modular exponentiation is implemented across the control and the work
registers by a unitary transform $U_{f}$, before which one eigenstate
of the work register is flipped (shown as the blue arrow) and after
which the eigenstates across the two registers are distinctly entangled
(shown by the black nodes). The two entangled states associated with
$\left|H\right\rangle $ and $\left|V\right\rangle $ directions in
each pulse are separately passed on to the Fourier transform stage,
respectively indicated by the solid and the dashed lines.~\label{Fig:circuit diagram}}
\end{figure}

To establish the state of Eq.~(\ref{eq: Shor entanglement}) in a
classical beam of light, we store the control register information
$\psi_{C}$ into the OAM-modes of an LG bream and work register information
$\psi_{W}$ into the polarization directions $\mathbf{H}$ and $\mathbf{V}$
of that beam. Hence, the initial separable input state can be written
as
\begin{equation}
\left|\psi_{\mathrm{in}}\right\rangle =\sum_{l=1}^{k}\left[\left|+l\right\rangle +\left|-l\right\rangle \right]\otimes\left|H\right\rangle ,\label{eq: input state}
\end{equation}
where the normalization constant is omitted to simplify the expression.
For an $n$-bit long integer $N$, $k$ should at least be equal to
$2^{n-1}$. Then, all algorithmic steps involved for computing the
multiplicative order described by the originally quantum algorithm
can be realized on Eq.~(\ref{eq: input state}) using just classical
light path manipulations. Fig.~\ref{Fig:circuit diagram} shows the
partitioning of the four major steps of Shor's algorithm: (i) preparing
the initial state; (ii) performing modular exponentiation (MEF); (iii)
performing quantum or discrete Fourier transform (DFT); and (iv) executing
read out. 

Before describing the detailed setup of an integrated light path for
carrying out the forementioned four steps, we examine from a formal
perspective how Eq.~(\ref{eq: input state}) changes in each stage
to demonstrate that the final state of the LG beam contains computationally
significant information. First, one notes the work register is limited
to two states $\left|H\right\rangle $ and $\left|V\right\rangle $.
To resolve this scaling limitation, we appeal to a pulse train implementation
where the number of pulses required to encode the modulus of the control
register is, in the worst case scenario where $N$ is prime, $n$.
On average, if a factor exists for $N$, only $n-1$ bits or pulses
are needed. In the meanwhile, the extention in pulse number allows
the control register information be distributed among multiple OAM
modes across pulses. Therefore, in general, one encodes the state
\begin{equation}
\left|\psi\right\rangle =\sum_{l=1}^{k}\prod_{i=1}^{\otimes n}\left[\left|+l_{i},H_{i}\right\rangle +\left|-l_{i},V_{i}\right\rangle \right],\label{eq:pulse_state}
\end{equation}
in an $n$-pulse train.

\begin{figure}[H]
\centering{}\includegraphics[bb=0bp 0bp 629bp 461bp,clip,width=10cm]{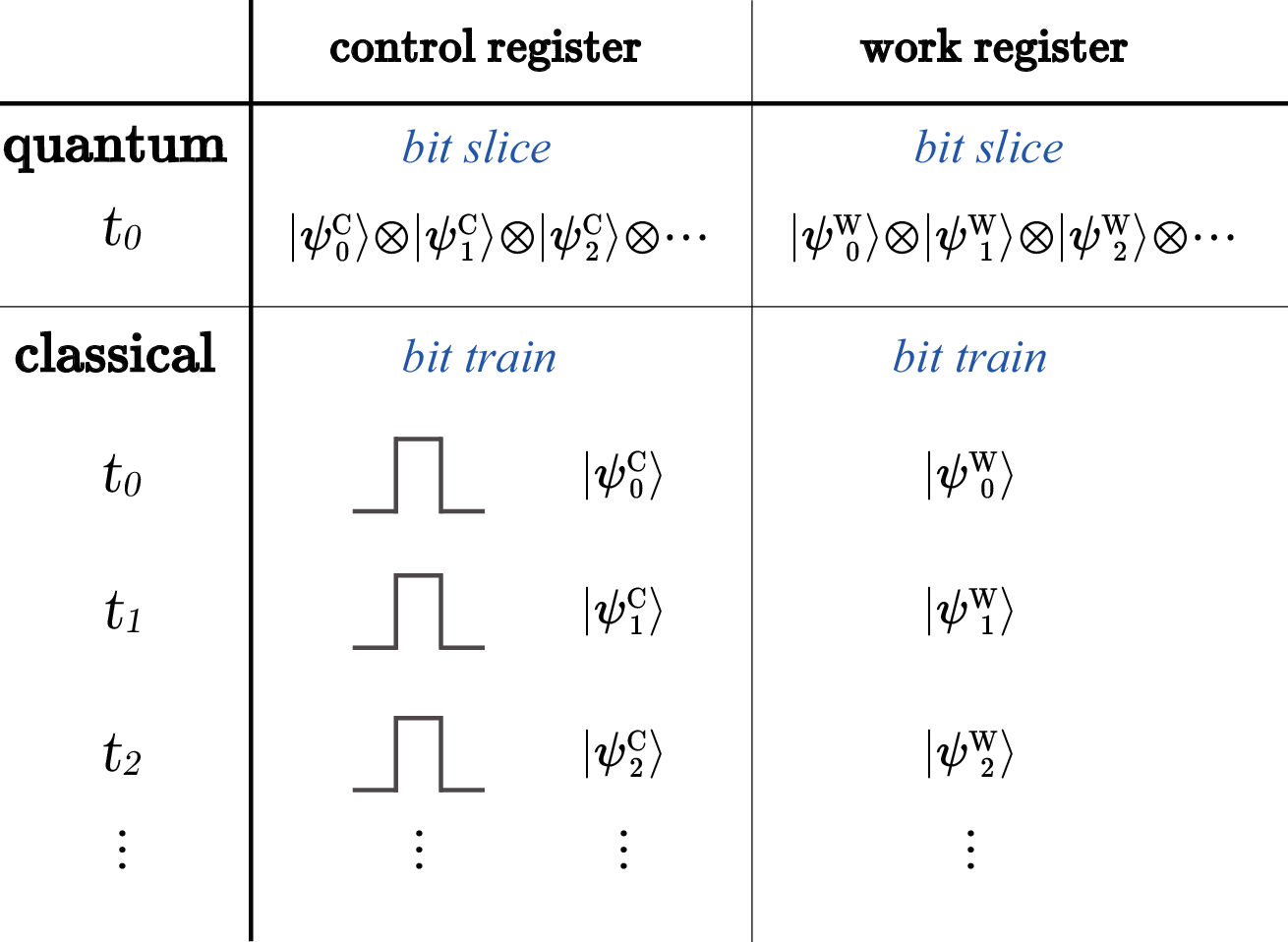}\caption{Illustration for the distinct data encoding schemes
used by two entangled registers in typical quantum entangling implementations
and our classical entangling implementation. In the former, the qubits
of the control register are collectively entangled with the qubits
of the work register; the bit-slice data in the two registers are
entangled as cluster entangled states. In the latter, since the pulses
are time ordered without inter-pulse interactions, the entangling
encoding extends in time and is done one bit in the control register
with one bit in the work register at a time. The time ordering is
equivalent to spatial ordering in a light space and does not contribute
to extra time complexity of the algorithm.~\label{Fig: different encoding}}
\end{figure}

Note that the encoding scheme of Eq.~(\ref{eq:pulse_state})
is different from typical encoding schemes used for implementing Shor's
algorithms. As illustrated in Fig.~\ref{Fig: different encoding},
the bits in both the control and the work registers are temporally
ordered to comply with the time-ordered optical pulses in a pulse
train. For the simplest case where two OAM modes are used to entangled
with the polarization directions, one pulse encodes the entangled
data of exactly one bit in $\left|\psi_{C}\right\rangle $ and one
bit in $\left|\psi_{W}\right\rangle $. The pulse train extends in
time to encode the multi-bit data necessary for both registers. In
contrast, conventional approaches on solid-state qubits and photon
pairs have the scaling done across space to encode entangled bit slices
collectively. One bit-slice data is encoded as a cluster state in
$\left|\psi_{C}\right\rangle $ to entangle with a cluster state in
$\left|\psi_{W}\right\rangle $ containing the data of an associated
bit slice. This distinction between vertical extension and horizontal
scaling affects the detection scheme at eventual readout. The quantum
projection or collapse measurement is conducted, theoretically, in
one shot, whereas the pulse train scheme requires bit-by-bit detection
on successive pulses. However, pulse detection on photo diodes occupies
fixed time. Algorithmically speaking, the complete decoding of a pulse
train is completed within one iteration; the extra complexity incurred
is the constant $\mathcal{O}(1)$. Therefore, the new encoding scheme
does not sacrifice extra algorithmic complexity to the factoring procedure.

The modular exponentiation $a^{x}\mod N$ establishes a specific correlation
between the computational bases in the control and work registers.
For example, if the OAM modes are limited to $\left\{ +1,-1\right\} $,
the register encoding becomes fully binary and one should have
\begin{equation}
\left|\psi_{\mathrm{MEF}}\right\rangle =\sum_{j=1}^{n}\left|x_{j},a^{x_{j}}\mod N\right\rangle \label{eq:MEF}
\end{equation}
after the exponentiation step, where $x_{j}$ indicates a binary sequence
encoded by the two OAM modes. 

For the case $N=15$, the MEF should set work register to 1 ($\left|\psi_{W}\right\rangle =\left|H\right\rangle $)
that associates with either the case $x=0$ or $x=2$, i.e.$\left|\psi_{C}\right\rangle =\left|+1\right\rangle $
or $\left|+2\right\rangle $, at the control register; complementarily,
the work register stores 11 ($\left|\psi_{W}\right\rangle =\left|V\right\rangle $)
for $x=1$ or $x=3$ at the control register, i.e. $\left|\psi_{C}\right\rangle =\left|-1\right\rangle $
or $\left|-2\right\rangle $. The entangled state after the MEF should
read

\begin{equation}
\sum_{l=1}^{2}\left[\left|+l,H\right\rangle +\left|-l,V\right\rangle \right],\label{eq: after MEF}
\end{equation}
where the necessary information for eliciting the correct multiplicative
order is contained. Therefore, using four OAM modes, one pulse from
the source and a CW laser is sufficient to encode for $N=15$.

The next step in Shor's original routine involves a projective readout
on Eq.~(\ref{eq: after MEF}) that collapses the work register into
a definite $\left|H\right\rangle $ or $\left|V\right\rangle $ state,
leaving the control register retain an eigenstate superposition. Note
that the MEF entanglement guarantees that either way of the collapse
will impose the same multiplicative order in the superposition. In
contrast, deprived of the unique property of quantum collapse, classical
beams can substitute this step by simply splitting the mixed beam
through a polarizing beam splitter (PBS). That is, the PBS would split
the pulse sequence of Eq.~(\ref{eq:MEF}) into two, each having the
same polarization direction among every pulse, e.g. one would have
$\left|\psi_{\mathrm{COL}}\right\rangle =\sum_{r}\left|x_{r},\mathcal{P}\right\rangle $
where \emph{$\left|\mathcal{P}\right\rangle $} is either $\left|H\right\rangle $
or $\left|V\right\rangle $ and $x_{r}=qr+m$ ($m$ is the modulus
and $q$ is the quotient).

In the $N=15$ example, collapsing $\left|\psi_{W}\right\rangle $
into either $\left|H\right\rangle $ or $\left|V\right\rangle $ will
retain identically two eigenstates in $\left|\psi_{C}\right\rangle $;
whether either $\left|+1\right\rangle +\left|+2\right\rangle $ or
$\left|-1\right\rangle +\left|-2\right\rangle $ points to the same
order 2 (i.e. $11^{2}\mod15=1$). 

\begin{align}
\left|\psi_{H}\right\rangle  & =\left|+1,H\right\rangle +\left|+2,H\right\rangle ,\nonumber \\
\left|\psi_{V}\right\rangle  & =\left|-1,V\right\rangle +\left|-2,V\right\rangle ,\label{eq: after tracing out}
\end{align}
on either of which the subsequent algorithmic steps can operate to
achieve integer factoring of $N$.

One of the polarized beam in Eq.~(\ref{eq: after tracing out}) then
undergoes the discrete Fourier transform (DFT) (the third major key
illustrated in Fig.~\ref{Fig:circuit diagram}) expressed as

\begin{equation}
\left|x\right\rangle \rightarrow\mathop{\frac{1}{\sqrt{2^{n}}}\sum_{j=0}^{2^{n}-1}}e^{i2\pi jx/2^{n}}\left|j\right\rangle .\label{eq: DFT_tfm}
\end{equation}
Each state $\left|x\right\rangle $ is transformed to a set of new
orthogonal bases $\left|j\right\rangle (j\in\{0,1,\dots,2^{n}-1\})$
with a specific distribution of phases determined by the value of
$x$ in $2^{n}$ dimensions. For typical quantum optical~\citep{Politi,Martin-lopez}
or superconducting circuit~\citep{Lucero} implementations, Hadamard
gates or control phase gates are consecutively applied to individual
qubits to realize the Fourier transform, making the transform of Eq.~(\ref{eq: DFT_tfm})
essentially an $n$-step one-qubit operation. The OAM space of the
LG beam considered here need not be decomposed into $2$-level spaces,
so the phases can be imposed concurrently on the OAM modes $\left|l\right\rangle $,
i.e. the DFT transform bases $\left|j\right\rangle $, once they are
separated. In the illustrated example, since $\left|j\right\rangle $
ranges over $\left|+1\right\rangle $, $\left|-1\right\rangle $,
$\left|+2\right\rangle $ and $\left|-2\right\rangle $, the intermediate
state resulted from $\left|l=-1\right\rangle $ for example becomes
\begin{equation}
\left|\psi_{\mathrm{int}}^{(-1)}\right\rangle =\left|+1\right\rangle +i\left|-1\right\rangle -\left|+2\right\rangle -i\left|-2\right\rangle .\label{eq:DFT_intermed}
\end{equation}
The states with phases $\{0,\pi/2,\pi,3\pi/2\}$ for different $j$
imposed can be realized by spiral phase plates (SPP) and phase-only
spatial light modulators (SLM) as well as q-plates~\citep{Marrucci}
and digital micromirror devices (DMD)~\citep{Mirhosseini}. While
the former performs the function of converting OAM modes, e.g. from
$\left|+1\right\rangle $ to $\left|+2\right\rangle $~\citep{Beijersbergen},
the latter introduces an arbitrary phase to a specific mode. Combining
the two allows one to generate $i\left|+1\right\rangle $ from $\left|-1\right\rangle $
through the addition of a $\pi/2$ phase~\citep{Song}. To complete
the DFT transform, each state that underwent Eq.~(\ref{eq: DFT_tfm})
should be combined through addition, i.e. to arrive at 
\begin{equation}
\left|\psi_{\mathrm{DFT}}\right\rangle =\left|\psi_{\mathrm{int}}^{(\pm1)}\right\rangle +\left|\psi_{\mathrm{int}}^{(\pm2)}\right\rangle ,\label{eq:DFT_result}
\end{equation}
where the sign depends on which beams we choose from Eq.~(\ref{eq: after tracing out}). 

For the demonstration case $N=15$, where the choice of the sign would
lead the control register to retain either the superposition $\left|+1\right\rangle +\left|+2\right\rangle $
or $\left|+1\right\rangle -\left|+2\right\rangle $. The phase difference
between the two basis vectors does not affect the computation significance.
Either state contains the OAM modes $l=+1$ and $l=+2$, which translate
into the computational values $0$ and $2$, respectively, according
to the algorithmic premise we have assumed. Since only a non-zero
value is acceptable, one is left with $x=2$ from which the multiplicative
order $r=2^{n}/x$ can be obtained. Then, the integer factors become
apparent by finding the greatest common divisor of $11^{r/2}\pm1$
and $N$. The last two conversion steps are not computationally complex
and thus not necessarily executed in the light paths.

\section{Optical paths for algorithmic steps\label{sec:optical_paths}}

The two major algorithmic steps according to Fig.~\ref{Fig:circuit diagram}
and the discussion in the last section are: (i) modular exponentiation
function (MEF) and (ii) discrete Fourier transform (DFT). Below we
examine the details of the experimental setup that can accomplish
these two steps.

\subsection{Modular exponentiation}

\begin{figure}[H]
\noindent \centering{}\includegraphics[clip,width=10cm]{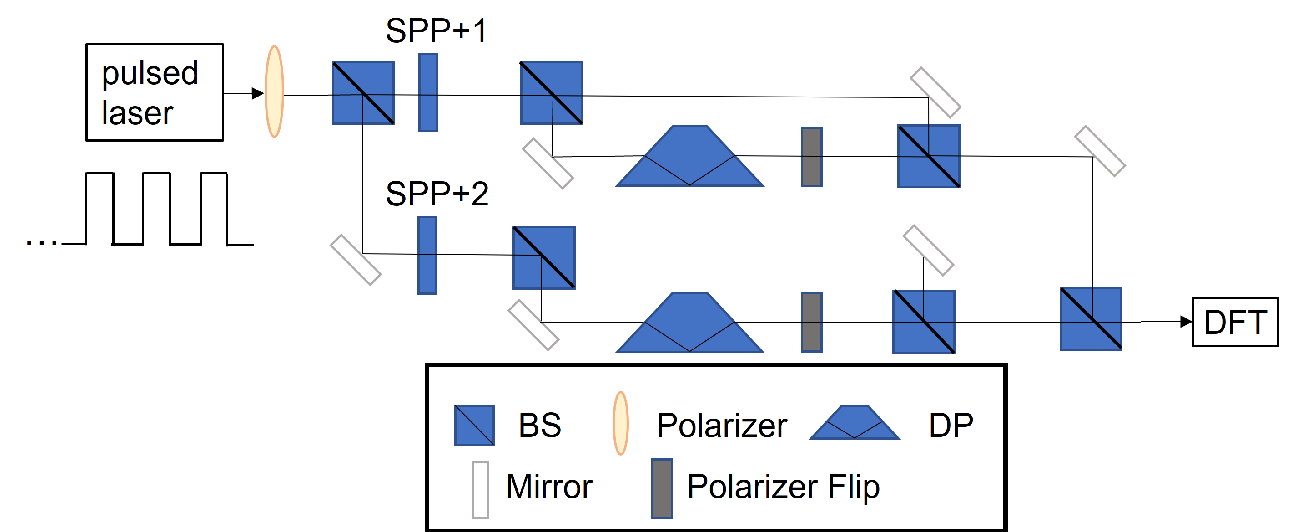}
\caption{Optical path diagram of modular exponentiation in Shor's algorithm.
A laser pulse train passes through a polarizer and SPPs to generate
the OAM modes of Laguerre-Gaussian beams, which refer to different
states in the algorithm. The DPs, polarizer flips, and BSs are used
to execute the algorithm. BS indicates 50/50 beam splitters, SPP spiral
phase plates, and DP Dove prisms.~\label{Fig:MEF}}
\end{figure}

The optical path setup for modular exponentiation is shown in Fig.~\ref{Fig:MEF}.
The light source is a phase-locked laser pulse train, which is let
pass through a horizontal polarizer and split into two equal intensity
branches of the same horizontal polarization along two paths. These
two paths are identical except for the spiral phase plates (SPP) which
modulate identical beams into distinct OAM modes. For the exemplifying
$N=15$ case, one could consider a train of one single pulse or, more
simply, a CW laser as the source and have OAM modes of $l=+1$ and
$l=+2$~\citep{Beijersbergen} representing the computational states
$\left|+1,H\right\rangle $ and $\left|+2,H\right\rangle $, respectively.
The beam along each path then passes through a beam splitter (BS),
from which one branch goes through a Dove prism (DP) to invert the
OAM sign ($+1\to-1$ or $+2\to-2$)~\citep{Gonz=0000E1lez} and a
polarizer set to flip $\left|H\right\rangle $ to $\left|V\right\rangle $,
while the other branch remains unchanged. Eventually, the four branches
of the two beams are recombined using three beam splitters, giving
rise to the desired state after MEF according to Eq.~(\ref{eq: after MEF}).

\subsection{Discrete Fourier transform}

\begin{figure}[H]
\noindent \centering{}\includegraphics[clip,width=10cm]{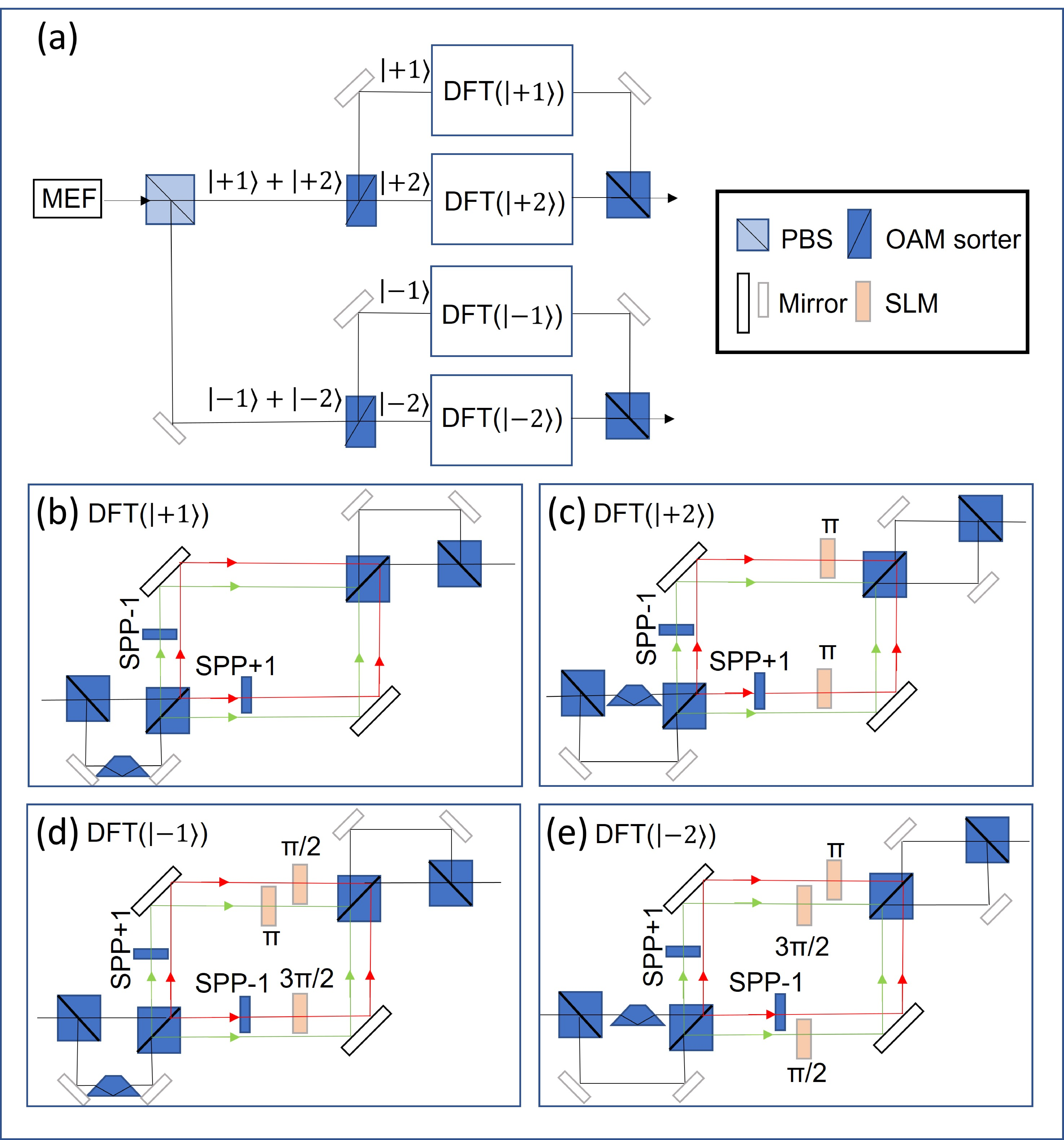}\caption{Optical path diagram of DFT in Shor's algorithm. The work register
is measured to different polarization directions by a PBS. Each beam
is sorted by an OAM sorter into two beams with different OAM and then
emulates the DFT individually. Finally combining the beams and getting
the output related to the order. SLM is phase-only spatial light modulator.
\label{Fig:DFT}}
\end{figure}

As explained in the last section, the classical light path needs to
achieve an effect analogous to quantum collapse to elicit pulse states
useful for computational purposes before feeding them into subsequent
stages for discrete Fourier transforms. To do this, the output from
the MEF stage is split into two through a polarizer, allowing one
to obtain two pulse trains, each carrying the same polarization direction
for all pulses whose OAM modes are either $\{\left|+1\right\rangle ,\left|+2\right\rangle ,\dots\}$
or $\{\left|-1\right\rangle ,\left|-2\right\rangle ,\dots\}$. Each
train is further split by an OAM sorter~\citep{Leach} to obtain
individual pulses of particular $\left|\pm j\right\rangle $. The
DFT process is effectively a transformation that imposes specific
phases on each pulse. Realized by phase gates on photonic systems,
it is in contrast realizable using Sagnac interferometry here for
the classical light beams. Each pulse corresponding to a particular
$\left|j\right\rangle $ is first superposed with its inverse state
$\left|-j\right\rangle $ through beam splitters and a Dove prism
before it enters its designated interferometer, in which $\left|j\right\rangle $
generates a superposition of all basis states of the same sign. 

For a concrete illustration, consider the optical path setup shown
in Fig.~\ref{Fig:DFT} for $N=15$ with the input MEF end being the
states given in Eq.~(\ref{eq: after tracing out}). The beam is first
polarizingly split before fed into OAM sorters, producing in total
four branches of distinct states, i.e. $\left|+1\right\rangle $,
$\left|-1\right\rangle $, $\left|+2\right\rangle $, and $\left|-2\right\rangle $.
Considering all these states running in parallel, DFT effectively
implements a $4\times4$ matrix transformation on them, where the
matrix entries represent complex phase coefficients of a Fourier transform.
To realize the resulting state in Eq.~(\ref{eq:DFT_result}), the
Sagnac interferometers, shown as boxes of Fig.~\ref{Fig:DFT}, use
SPPs to generate new OAM basis, e.g. $\left|+1\right\rangle \to\left|+1\right\rangle +\left|+2\right\rangle $
and $\left|-1\right\rangle \to\left|-1\right\rangle +\left|-2\right\rangle $,
whilst using the SLMs subsequently to generate corresponding phases
along the state-designated paths. The beams are eventually combined
using BS to output the computed state as in Eq.~(\ref{eq:DFT_result}).
This state then enters the final stage of intereference fringe detection
to obtain the information of multiplicative order.

\section{Detection method and results\label{sec:detection_method}}

\begin{figure}[H]
\noindent \centering{}\includegraphics[width=9cm]{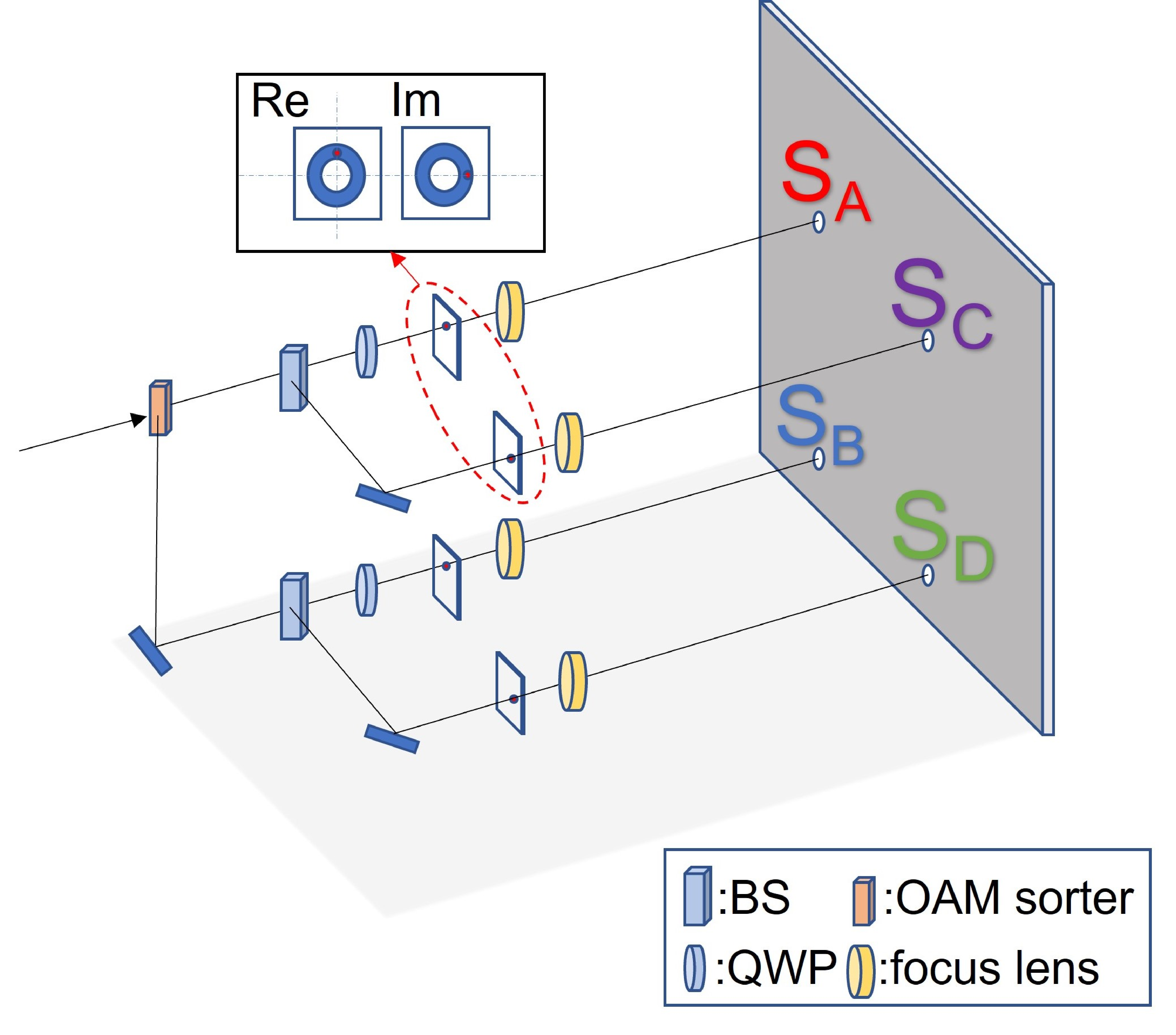}\caption{Optical path diagram of the detection method. The DFT output beam
first goes through a OAM sorter to separate the beam into different
LG modes. Then two beams are decomposed into four beams by two BSs.
By the orthogonal aperture diaphragms, the LG beams are separated
into the real and imaginary components. Finally, refocused beams enter
the four holes as the interference sources. The elements in the upper
box represent the orthogonal aperture diaphragms, BS is beam splitter,
QWP is quarter-wave plate. \label{Fig:detection path diagram}}
\end{figure}

The DFT stage outputs a train of pulses containing different OAM states
with computational values. The last step of the setup is light paths
that are responsible for telling what computational states are contained
in these LG beams. To visually discern the computational states, we
follow a proposal given in Ref.~\citep{You} to let the distinct
OAM states interfere with each other, generating unique fringe patterns
that are one-one corresponding to the quantum states. In other words,
the desired multiplicative order is implied in a two-dimensional fringe
image projected on a screen, as illustrated in Fig.~\ref{Fig:detection path diagram}.
Having a pulse train to generate the interference implies the fringe
patterns be recorded over multiple frames, where the frame rate corresponds
to the pulse width and the pulse period. Since for $N=15$, one frame
is sufficient, we illustrate the interference generation below with
a CW laser source in mind.

To implement the interference apparatus, the output from DFT is fed
into an OAM sorter to obtain two beams with different $l$ modes.
Then along each beam path, we use a BS to further split it into two
equal parts, constituting four beam paths. Two quarter-wave plates
(QWP) are used to compensate for the phase shift introduced by BSs.
One orthogonal aperture diaphragm is placed at each transmitted path
to select two points from the real and imaginary components of each
OAM state. For example, for the LG beam in Eq.~\ref{eq: LG equation},
the intensity distribution is a helical doughnut with a phase dependence
of exp$(-il\varphi)$. The choice of real and imaginary parts depends
on the azimuth $\varphi$, which is shown in the illustration of Fig.~\ref{Fig:detection path diagram}.
Two points at the vertical and horizontal axis of the intensity distribution
of two equal beams are chosen as the real and imaginary parts, respectively.
Finally, the deflected beams from the holes are refocused, acting
as the four sources $S_{A}$, $S_{B}$, $S_{C}$ and $S_{D}$ for
the four-hole interference. For better interference effect, motorized
translation stages (not drawn in the figure) can be used to synchronize
the delays among paths.

\begin{figure}[H]
\noindent \begin{centering}
\includegraphics[bb=0bp 10bp 366bp 310bp,clip,width=10cm]{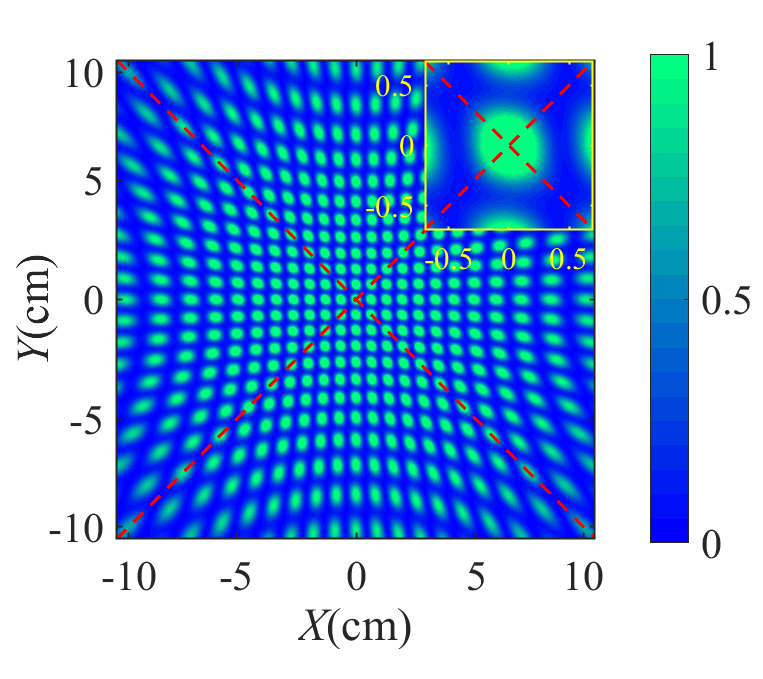}
\par\end{centering}
\caption{Interference patterns after DFT for state $\left|+1\right\rangle $$+$$\left|+2\right\rangle $
in the case $N=15$. $\left|+1\right\rangle $ and $\left|+2\right\rangle $
means $0$ and $2$ respectively. $0$ is obviously invalid. It means
order $r=2$. \label{DFT_result1}}
\end{figure}

\begin{figure}[H]
\noindent \centering{}\includegraphics[bb=0bp 10bp 366bp 310bp,clip,width=10cm]{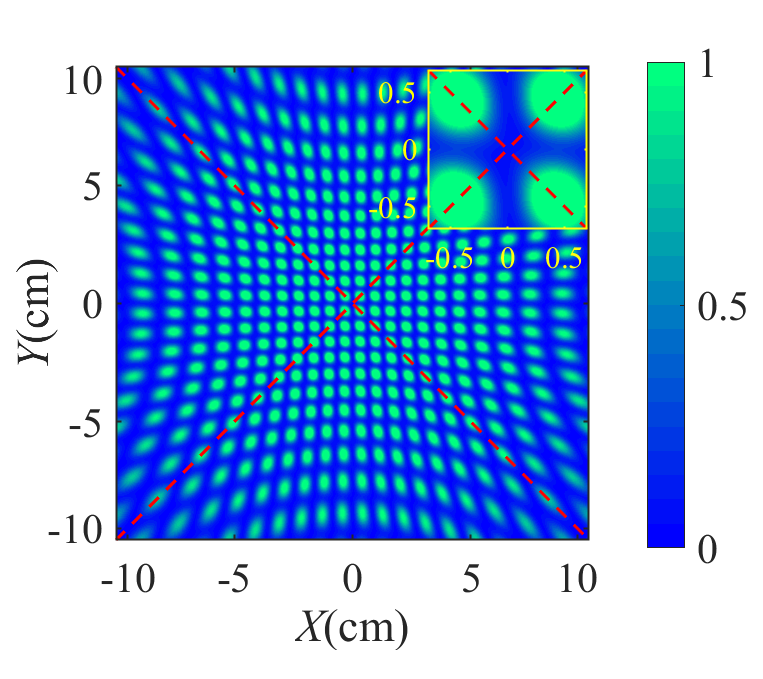}\caption{Interference patterns after DFT for state $\left|+1\right\rangle $$-$$\left|+2\right\rangle $
in the case $N=15$. It also means order $r=2$. \label{DFT_result2}}
\end{figure}

We simulate the interference patterns by encoding the four light sources.
A source laser wavelength of $632$~nm, a hole to hole distance of
$10\;\mu$m, a hole to screen distance of 10~cm, and a screen area
of $10\times10\mathrm{cm^{2}}$ are assumed. Meanwhile, we set the
$z=0,p=0$ in Eq.~(\ref{eq: LG equation}) for the source beam for
simplification. The interference fringe patterns are shown in Figs.~\ref{DFT_result1}
and \ref{DFT_result2}.

The pattern in Fig~\ref{DFT_result1} indicates the first result
for the state of Eq.~\ref{eq:DFT_result} after DFT in upper path
of Fig~\ref{Fig:DFT}(a), which is $\left|+1\right\rangle $$+$$\left|+2\right\rangle $.
We can see that the pattern is centrosymmetric, where the central
area is a bright spot. With these special properties, we can identify
the corresponding states in specific experiments. According to the
definition of DFT bases in Eq.~\ref{eq: DFT_tfm}, $\left|+1\right\rangle $
and $\left|+2\right\rangle $ refer to $0$ and $2$, respectively.
The order cannot be $0$, so if the pattern in Fig~\ref{DFT_result1}
appears in the experiment, we can extract effective information, that
is, order $r=2$.

Fig.~\ref{DFT_result2} shows another case in lower path of Fig.~\ref{Fig:DFT}(a)
after DFT. It refers to the result $\left|+1\right\rangle -\left|+2\right\rangle $.
The resulting fringe is still centrosymmetric but with a dark area
in the center. This pattern also corresponds to order $r=2$. As opposed
to quantum detection, our method can detect the different states in
Eq.~\ref{eq:DFT_result} simultaneously only if we connect a detection
system behind the multiple output optical paths in Fig~\ref{Fig:DFT}.
This is because that the probabilistic solution in the quantum implementation
can be equivalent to the probability of choosing one or more of the
outputs. Therefore, we can determine the reliability of the results
by detecting multiple outputs. 

\section{Complexity analysis\label{sec:Complexity-analysis}}

We compare the complexity of the best classical number field sieve
algorithm and the original Shor's algorithm for factoring $N$ with
our method in the Table~\ref{Complexity}. Firstly, It has been proven
that both the memory space and time of number field sieve is exponential
with $N$.~\citep{Bernstein,Anand} This means that the computing
resources and time required will increase significantly as $N$ increases.
While quantum bits with supersition and entanglement properities are
used to decompose large numbers in polynomial time in original Shor's
algorithm. Basically, It needs $3\lceil\textrm{log}_{2}N\rceil$ qubits
in two registers and $O((\textrm{log}N){}^{2}(\textrm{loglog}N)(\textrm{logloglog}N))$
steps.~\citep{Shor2,Beckman}

In contrast, our method only needs $\lceil\textrm{log}_{2}N\rceil$
pulses in Eq.~(\ref{eq:pulse_state}) with the advantange of OAM
mode and time degree of freedom. In each pulse, we can use the OAM
modes and polarization directions to encode two registers respectively.
Also, these two degrees of freedom in different pulses can form a
binary sequence in a extensible version. For the equivalent time complexity
in our method, the MEF adopts the direct encoding for the function
result obtained from the external computing platform such as a field-programmable
gate array (FPGA). So the time complexity of MEF is similar to the
original Shor's algorithm. Considering the time complexity of each
optical element as $O(1)$, the DFT we used in the algorithmic step
has the time complexity $O(\textrm{log}N)$ with the advantage of
parallel computing in Fig.~\ref{Fig:DFT}. Therefore the total time
complexity is still consistent with the original Shor's algorithm.
In brief, our macroscopic algorithmic steps using classical light
achieve the same complexity as Shor's algorithm on the basis of ensuring
space consistency in order of magnitude. 
\begin{table}[H]
\begin{centering}
\begin{tabular}{|c|c|c|c|}
\hline 
 & Memory space  & Time complexity  & Reference\tabularnewline
\hline 
\hline 
Number field sieve  & $\textrm{exp}\left\{ \frac{c}{2}(\textrm{log}N)^{1/3}(\textrm{log}\textrm{log}N)^{2/3}\right\} $
bits  & $O(\textrm{exp}\left\{ c(\textrm{log}N)^{1/3}(\textrm{log}\textrm{log}N)^{2/3}\right\} )$  & \citep{Bernstein,Anand}\tabularnewline
\hline 
Original Shor's algorithm & $3\lceil\textrm{log}_{2}N\rceil$ qubits & $O((\textrm{log}N){}^{2}(\textrm{loglog}N)(\textrm{logloglog}N))$  & \citep{Shor2,Beckman}\tabularnewline
\hline 
Our method & $\lceil\textrm{log}_{2}N\rceil$ pulses & $O((\textrm{log}N){}^{2}(\textrm{loglog}N)(\textrm{logloglog}N))$ & this work\tabularnewline
\hline 
\end{tabular}
\par\end{centering}
\centering{}\caption{Complexity comparison of number field sieve, Shor's algorithm and
our method for factoring $N$.\label{Complexity}}
\end{table}

\section{Conclusions\label{sec:conclusions}}

In summary, we have implemented the major computational steps in Shor’s
algorithm using a classical optical system. The classical entangled
pair of OAM mode and polarization in LG beams or pulse trains are
used as the control and the work registers, respectively. A detection
method using four-hole interference for the entangled states is demonstrated
to obtain the multiplicative order $r$, which is equivalent to having
found a factor once Euclid's classical gcd algorithm is applied. Sourcing
the interfering light beams from fixed locations of the beam profiles,
we have shown that the multiplicative order is derivable from the
unique fringe patterns.

Compared to its quantum counterpart, the classical
approach avoids the implementations of time-based quantum gates while
maintain the same time complexity. The conditions for the generation,
operation, and detection of milliwatt-level light pulses are much
less stringent, compared to the implementations using single photon
pairs, which have low generation and detection efficiency. The former
is therefore less expensive to implement and more scalable. In addition,
the coherent laser beams is less prompt to the decoherence and noisy
environments suffered by superconducting circuits. Nevertheless, since
the data are encoded in the time-ordered pulses, the detection through
the interference scheme requires the effort to decode the information
from complex 2-dimensional images. Hence, future works would be devoted
to the simplification of the detection methodology.

From an information and computational perspective, the work here extends
the boundary of classical optics into the quantum regime. Like the
other works on emulating quantum behavior with classical optics systems,
it shows the line separating the quantum from the classical world
is less defined as physicists usually thought.

\section*{Data availability}
The datasets used and/or analyzed during the current study are available from the corresponding author on reasonable request.

\section*{Acknowledgments}

H.I. thanks the support by FDCT of Macau under grants 0130/2019/A3
and 0015/2021/AGJ and by University of Macau under grant MYRG2018-00088-IAPME.


\begin{thebibliography}{99}
\bibitem{Einstein}Einstein, A., Podolsky, B. \& Rosen, N. ``Can Quantum-Mechanical
Description of Physical Reality Be Considered Complete?'' Phys. Rev.
\textbf{47}, 777--780 (1935). 

\bibitem{Bell}Bell, J. S. ``On the Einstein Podolsky Rosen paradox,''
Physics Physique Fizika \textbf{1}, 195--200 (1964). 

\bibitem{Clauser}Clauser, J. F., Horne, M. A., Shimony, A. \& Holt,
R. A. ``Proposed Experiment to Test Local Hidden-Variable Theories,''
Phys. Rev. Lett. \textbf{23}, 880--884 (1969). 

\bibitem{Mermin}Mermin, N. D. ``Extreme quantum entanglement in a
superposition of macroscopically distinct states,'' Phys. Rev. Lett.
\textbf{65}, 1838--1840 (1990). 

\bibitem{Horodecki}Horodecki, R., Horodecki, P., Horodecki, M. \&
Horodecki, K. ``Quantum entanglement,'' Rev. Mod. Phys. \textbf{81},
865--942 (2009).

\bibitem{Pang}Pang, J.-Y. \& Chen, J.-W. ``On the renormalization
of entanglement entropy,'' AAPPS Bulletin \textbf{31}, 1-7 (2021).

\bibitem{Hobson}Hobson, A. ``Entanglement and the Measurement Problem,''
Quantum Engineering \textbf{2022}, 5889159 (2022).

\bibitem{Spreeuw}Spreeuw, R. J. C. ``A Classical Analogy of Entanglement,''
Found. Phys. \textbf{28}, 361--374 (1998). 

\bibitem{Qian}Qian, X.-F. \& Eberly, J. H. ``Entanglement and classical
polarization states,'' Opt. Lett. \textbf{36}, 4110--4112 (2011). 

\bibitem{Aiello}Aiello, A., Töppel, F., Marquardt, C., Giacobino,
E. \& Leuchs, G. ``Quantum-like nonseparable structures in optical
beams,'' New J. Phys. \textbf{17}, 043024 (2015). 

\bibitem{Qian-1}Qian, X.-F., Little, B., Howell, J. C. \& Eberly,
J. H. ``Shifting the quantum-classical boundary: theory and experiment
for statistically classical optical fields,'' Optica \textbf{2}, 611--615
(2015). 

\bibitem{Song}Song, X., Sun, Y., Li, Y., Qin, H. \& Zhang, X. ``Bell's
measure and implementing quantum Fourier transform with orbital angular
momentum of classical light,'' Sci. Rep. \textbf{5}, 14113 (2015). 

\bibitem{Shor}Shor, P. W. ``Algorithms for quantum computation: discrete
logarithms and factoring,'' in Proceedings 35th Annual Symposium on
Foundations of Computer Science (1994), pp. 124--134.

\bibitem{Shor2}Shor, P. W. ``Polynomial-Time Algorithms for Prime
Factorization and Discrete Logarithms on a Quantum Computer'', SIAM
J. Comput. \textbf{26}, 1484 (1997). 

\bibitem{Lucero}Lucero, E. \textit{et al}. ``Computing prime factors
with a Josephson phase qubit quantum processor,'' Nat. Phys. \textbf{8},
719--723 (2012).

\bibitem{Vandersypen}Vandersypen, L. M. K. \textit{et al}. ``Experimental
realization of Shor’s quantum factoring algorithm using nuclear magnetic
resonance,'' Nature \textbf{414}, 883--887 (2001). 

\bibitem{O=002019Brien}O’Brien, J. L. ``Optical Quantum Computing,\textquotedbl{}
Science \textbf{318}, 1567--1570 (2007). 

\bibitem{Lu}Lu, C.-Y., Browne, D. E., Yang, T. \& Pan, J.-W. ``Demonstration
of a Compiled Version of Shor’s Quantum Factoring Algorithm Using
Photonic Qubits,'' Phys. Rev. Lett. \textbf{99}, 250504 (2007). 

\bibitem{Lanyon}Lanyon, B. P. \textit{et al}. ``Experimental demonstration
of a compiled version of Shor's algorithm with quantum entanglement,\textquotedbl{}
Phys. Rev. Lett. \textbf{99}, 250505 (2007).

\bibitem{Politi}Politi, A., Matthews, J. C. F. \& O’Brien, J. L.
``Shor’s Quantum Factoring Algorithm on a Photonic Chip,'' Science
\textbf{325}, 1221--1221 (2009). 

\bibitem{Martin-lopez}Martín-López, E. \textit{et al}. ``Experimental
Realization of Shor’s Quantum Factoring Algorithm Using Qubit Recycling'',
Nat. Photon. \textbf{6}, 773--776 (2012).

\bibitem{Forbes}Forbes, A., Aiello, A. \& Ndagano, B. ``Chapter Three
- Classically Entangled Light,'' in Progress in Optics, T. D. Visser,
ed. (Elsevier, 2019), 64, pp. 99--153. 

\bibitem{Ladd}Ladd, T. D. \textit{et al}. ``Quantum computers,''
Nature \textbf{464}, 45--53 (2010). 

\bibitem{T=0000F6ppel}Töppel, F., Aiello, A., Marquardt, C., Giacobino,
E. \& Leuchs, G. ``Classical entanglement in polarization metrology,''
New J. Phys. \textbf{16}, 073019 (2014). 

\bibitem{Spreeuw-1}Spreeuw, R. J. C. ``Classical wave-optics analogy
of quantum-information processing,'' Phys. Rev. A \textbf{63}, 062302
(2001). 

\bibitem{Deutsch}Deutsch, D. \& Penrose, R. ``Quantum theory, the
Church--Turing principle and the universal quantum computer,'' Proceedings
of the Royal Society of London. A. Mathematical and Physical Sciences
\textbf{400}, 97--117 (1985). 

\bibitem{Perez-Garcia}Perez-Garcia, B. \textit{et al}. ``Quantum
computation with classical light: The Deutsch Algorithm,'' Phys. Lett.
A \textbf{379}, 1675--1680 (2015). 

\bibitem{Goyal}Goyal, S. K., Roux, F. S., Forbes, A. \& Konrad, T.
``Implementing Quantum Walks Using Orbital Angular Momentum of Classical
Light,'' Phys. Rev. Lett. \textbf{110}, 263602 (2013).

\bibitem{You}You, Z., Wang, Y., Tang, Z. \& Ian, H. ``Measurement
of Classical Entanglement Using Interference Fringes,'' J. Opt. Soc.
Am. B \textbf{38}, 1798 (2021). 

\bibitem{Allen}Allen, L., Beijersbergen, M. W., Spreeuw, R. J. C.
\& Woerdman, J. P. ``Orbital angular momentum of light and the transformation
of Laguerre-Gaussian laser modes,'' Phys. Rev. A \textbf{45}, 8185--8189
(1992). 

\bibitem{Beijersbergen}Beijersbergen, M. W., Coerwinkel, R. P. C.,
Kristensen, M. \& Woerdman, J. P. ``Helical-wavefront laser beams
produced with a spiral phase plate,'' Opt. Commun. \textbf{112}, 321--327
(1994). 

\bibitem{Marrucci}Marrucci, L., Manzo, C. \& Paparo, D. ``Optical
spin-to-orbital angular momentum conversion in inhomogeneous anisotropic
media,'' Phys. Rev. Lett. \textbf{96}, 163905 (2006).

\bibitem{Mirhosseini}Mirhosseini, M. \textit{et al}. ``Rapid generation
of light beams carrying orbital angular momentum,\textquotedbl{} Opt.
Express, \textbf{21}, 30196--30203 (2013). 

\bibitem{Gonz=0000E1lez}González, N., Molina-Terriza, G. \& Torres,
J. P. ``How a Dove prism transforms the orbital angular momentum of
a light beam,'' Opt. Express, \textbf{14}, 9093--9102 (2006).

\bibitem{Leach}Leach, J., Padgett, M. J., Barnett, S. M., Franke-Arnold,
S. \& Courtial, J. ``Measuring the Orbital Angular Momentum of a Single
Photon,'' Phys. Rev. Lett. \textbf{88}, 257901 (2002). 

\bibitem{Bernstein}Bernstein, D. J. ``Circuits for Integer Factorization:
A Proposal,\textquotedbl{} (2001). 

\bibitem{Anand}Anand, C., Gungor, A. \& Thomas, K. A. ``Factoring
of large numbers using number field sieve-the matrix step,\textquotedbl (2007).

\bibitem{Beckman}Beckman, D., Chari, A. N., Devabhaktuni, S. \& Preskill,
J. ``Efficient networks for quantum factoring,\textquotedbl{} Phys.
Rev. A \textbf{54}, 1034--1063 (1996).
\end{thebibliography}
\end{document}